\documentclass[twocolumn,prl,superscriptaddress]{revtex4-1}
\usepackage{amsmath,amssymb,mathrsfs}
\usepackage{natbib}
\usepackage{subfigure}
\usepackage{tabularx}
\usepackage{epsfig}
\usepackage{longtable}
\usepackage{amsfonts}
\usepackage{rotating}
\usepackage{bbold}
\usepackage{hhline}
\usepackage{braket}
\usepackage{txfonts, comment}
\usepackage{multirow}

\usepackage{appendix}
\usepackage[unicode=true,bookmarks=true,bookmarksnumbered=false,bookmarksopen=false,breaklinks=false,pdfborder={0 0 1},backref=false,colorlinks=true]{hyperref}

\hypersetup{linkcolor=magenta,urlcolor=blue,citecolor=blue,pdfstartview={FitH},hyperfootnotes=false,unicode=true}

\def\be{\begin{equation}}
\def\ee{\end{equation}}
\def\bea{\begin{eqnarray}}
\def\eea{\end{eqnarray}}

\begin{document}

\title{The mixing-spacetime symmetry in the Floquet-Bloch band theory}

\author{Pei Wang}
\affiliation{Department of Physics, Zhejiang Normal University, Jinhua 321004, China}
\email{wangpei@zjnu.cn}

\begin{abstract}
We discover a class of spacetime symmetries
unique to time-periodic systems, which we term "mixing symmetry"
due to its combination of space and time coordinates
in the symmetry transformation. We systematically enumerate
the symmetry groups, and classify the corresponding Floquet-Bloch band theories
by utilizing the winding number of quasi-energy. Moreover,
we provide a comprehensive scheme for the experimental realization of these symmetries.
The particle propagator exhibits an intriguing pattern that remains invariant
even under transformations mixing space and time coordinates.
We anticipate that this distinct feature can be observed in current cold atom experiments.
\end{abstract}

\date{\today}

% Make the title.
\maketitle

{\it Introduction.---} 
The study of Floquet-Bloch bands has emerged as a central
topic in the field of nonequilibrium driven many-body
dynamics~\cite{Oka09,Kitagawa10,Lindner11,Cooper19,Rudner20}.
Recent advancements in precise control and probing
techniques have allowed for the realization of Floquet-Bloch
bands in diverse platforms, including photonic waveguides~\cite{Rechtsman13},
solid materials~\cite{Wang13}, and cold atom systems~\cite{Jotzu14}. By employing
periodic driving, these systems offer a unique opportunity to
explore models that are challenging to realize in static setups~\cite{Sorensen05,
Eckardt05, Goldman14,Oka19}.
Moreover, periodic driving enables the emergence of new
states of matter that lack a static analog, leading to
captivating phenomena in condensed matter physics, such as
symmetry breaking~\cite{Else16,Khemani16,Yao17},
localization~\cite{Ponte15,Lazarides15,Bordia17}, and
topological effects~\cite{Kitagawa10,Lindner11,Cooper19,Rudner20,Rudner13}.

Symmetry plays a fundamental role in the study of band theory,
exerting profound effects on various aspects of band structures.
Spatial symmetries such as rotation, mirror reflection, and
space inversion have long been recognized for their ability
to protect band crossings or generate degeneracies~\cite{Ashcroft76}.
Time reversal, particle-hole, and chiral symmetry have been
utilized in the renowned tenfold classification of
insulators and superconductors~\cite{Altland97}.
This classification has been applied to provide a periodic
table for the topological phases~\cite{Schnyder08,Kitaev09}, and more recently,
it has been extended to the topological classification of
Floquet-Bloch bands~\cite{Nathan15,Potter16,Else16b,Roy17}.
Moreover, researchers have recognized the critical interplay between
space group symmetries and topology,
culminating in the comprehensive topological classification of
band structures for all 230 crystal symmetry groups~\cite{Bradlyn17,Po17,Kruthoff17}.
Notably, in the realm of Floquet systems, the presence of
intertwined spatial and temporal translations, including
nonsymmorphic symmetries such as glide time-reversal
or time glide reflection, can preserve spectral degeneracy
and give rise to novel out-of-equilibrium phases~\cite{Morimoto17,Xu18,Peng19,Mochizuki20}.

But previous studies have overlooked a class of symmetries
that is unique to time periodic systems and absent in static ones.
These symmetries are referred to as mixing symmetries in this paper. Let us
consider the coordinates in 1+1-dimensional spacetime as
$\left(t,x\right)$. A linear coordinate transformation can be represented as
$\left(t',x'\right)^T=A \left(t,x\right)^T$. If the matrix $A$ contains
non-zero off-diagonal elements, it is called a mixing transformation
because it combines the space and time coordinates.
One well-known example of a mixing symmetry is the
Lorentz symmetry, which holds significant importance in
quantum field theory. In the context of condensed matter
physics, the Schr\"{o}dinger equation treats space and time
differently, making continuous mixing symmetry impossible.
Nevertheless, this does not rule out the possibility of discrete mixing symmetries~\cite{Wang18,Wang20}.
Recently, it has been discovered that the spacetime crystals that
exhibit a discrete Lorentz symmetry can be realized
in ultracold atomic gases confined to an optical lattice~\cite{Wang21,Wang22}.
But the models are constructed on finite-sized
lattices and do not exhibit continuous Floquet-Bloch bands
in the thermodynamic limit.

In this paper, we present the first evidence of the existence of
continuous Floquet-Bloch band theory that incorporates mixing
symmetry. We thoroughly identify and classify the
mixing groups in 1+1 dimensions. The resulting Floquet-Bloch
theories are categorized based on both symmetry groups and
the winding number of quasi-energy in the Brillouin zone (see Tab.~\ref{tab}
for a summary). Unlike previously studied symmetries, the operator
of mixing symmetry does not commute or anti-commute with the Hamiltonian.
Therefore, we rely on group representation theory for
constructing models. The band theory with mixing symmetry
exhibits a quasi-energy-momentum relation that remains
invariant under mixing transformations. Consequently, the particle
propagator in real spacetime exhibits invariance when the spacetime
coordinates undergo the transformation $A$,
which is a distinctive characteristic of mixing symmetry.

We discuss the possible realization of mixing symmetry in
cold atoms on an optical lattice. The precise
control achieved at the single-site level in experiments
allows for programmable Hamiltonians with locally adjustable
potential energies on each lattice site,  facilitated by
microelectromechanical systems mirrors~\cite{2016_Weiss_Science, browaeys2020many}.
We show that the Floquet-Bloch band with mixing symmetry can
be implemented using a quadratic quantum Fourier
transform (QQFT) protocol~\cite{Wang22} on a driven optical lattice that features
only onsite potential and nearest-neighbor hopping. The mixing symmetry can be
observed by locating a Bose-Einstein condensate
on the lattice and monitoring the atom density.

\begin{figure}[htp]
\centering
\vspace{0.2cm}
\includegraphics[width=.45\textwidth]{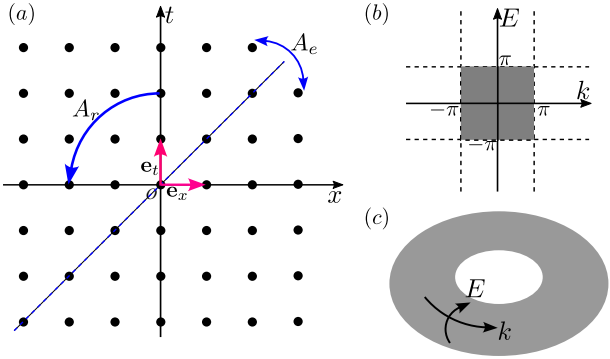}
\caption{(a) Schematic diagram of a spacetime crystal.
The primitive vectors ${\bf e}_t$ and ${\bf e}_x$ define the unit cell.
The mixing symmetry transformations, $A_e$ and $A_r$,
correspond to the exchange and rotation operations, respectively.
(b) and (c) depict the Floquet-Bloch-Brillouin zone and its topological equivalence.}
\label{fig:stlattice}
\end{figure}

\emph{Method.}---
When studying a quantum model, the usual approach involves
writing down the Hamiltonian in real spacetime and then
extracting the underlying symmetry from it. However,
our objective is to construct a model with specific symmetry.
We begin by providing a complete list of the
mixing symmetry groups. Next, we establish the unitary
representation of each group within the quasi-momentum-energy space. In this
representation, the symmetry manifests as a constraint
on the dispersion relation (DR), which is the function $E(k)$
describing the quasi-energy $E$ as a function of quasi-momentum $k$.
For continuous Floquet-Bloch bands, we discover a fundamental
equation governing the topology of the DR, which
serves as a basis for band classification. By finding an $E(k)$
that satisfies both the symmetry and topology conditions,
we obtain the Floquet Hamiltonian $\hat{H}_F$.
Finally, we demonstrate how to realize $\hat{H}_F$
using a time-periodic Hamiltonian $\hat{H}(t)$ with locality
in real spacetime. Our approach is inspired by the principles
of quantum field theory, which relies on the unitary
representation of the Poincar\'{e} group~\cite{Weinberg}.

\begin{table}[b]
\renewcommand\arraystretch{2.0}
\begin{tabular}{| c | c | c | c | c |}
\hline
& $A$ & $\bar{A}$  & Bands & Winding of DR\\
\hline
 \multirow{2}{*}{$\mathcal{P}_2$} & 
 \multirow{2}{*}{$\left(\begin{array}{cc} a & b \\ c & -a \end{array}\right)$} & \multirow{2}{*}{$-A$} &
 singlet &
 $w=\frac{-a-1}{b} \ \text{or} \ \frac{1-a}{b}$ \\
\cline{4-5} & & &
doublet & $w_1=w_2= \frac{-a-1}{b} \ \text{or} \ \frac{1-a}{b}$ \\
\hline
 \multirow{2}{*}{$\mathcal{P}_4$} & \multirow{2}{*}{$\left(\begin{array}{cc} a & b \\ c & -a \end{array}\right)$} 
  & \multirow{2}{*}{$A$} & doublet &
 $w_1=\frac{1-a}{b}, \ w_2 = \frac{-a-1}{b}$  \\
\cline{4-5} & & & quadruplet &
$w_1=w_3=\frac{1-a}{b}, \ w_2 = w_4=\frac{-a-1}{b}$ \\
\hline
\end{tabular}
\caption{Classification of Floquet-Bloch bands with
mixing symmetry. $A$ represents the coordinate transformation with
$bc = -a^2\pm 1$ for the symmetry classes $\mathcal{P}_2$ and $\mathcal{P}_4$,
respectively, while $\bar{A}$ corresponds to the transformation
in the $k$-$E$ space. The bands are categorized as singlets,
doublets, and quadruplets based on their mapping under
$\bar{A}$. The winding number, denoted as $w$, distinguishes
the winding properties of different bands.}\label{tab}
\end{table}

%The result is summarized in Table~\ref{tab}. $\mathcal{P}_2$
%and $\mathcal{P}_4$ are the only mixing groups in which there exist continuous Floquet-Bloch bands.
%$A$ is the symmetry transformation in real spacetime, and $\bar{A}$ is its
%dual transformation in $k$-$E$ space. Within each symmetry, the bands
%are classified according to their map under $\bar{A}$, and the DR winding numbers.

\emph{Mixing groups.}---
We are considering a 1+1-dimensional spacetime where
spatial rotation or mirror reflection is absent, allowing us
to concentrate on the study of mixing symmetry.
There are two noncollinear translational vectors.
Without loss of generality, we assign one vector
(${ \mathbf{e}_x}$) to the spatial direction (${x}$ axis),
and the other vector (${ \mathbf{e}_t}$) to the temporal
direction ($t$ axis), as shown in Fig.~\ref{fig:stlattice}.
This choice can always be made by employing a coordinate
transformation that rotates ${ \mathbf{e}_t}$ and
${ \mathbf{e}_x}$ into the $t$ and $x$ axes, respectively.
It is important to note that we exclusively consider
symmorphic groups in this paper. For nonsymmorphic groups,
nonsymmorphic symmetries may become significant when
${ \mathbf{e}_t}$ and ${ \mathbf{e}_x}$ are not orthogonal to each other~\cite{Xu18}.
To simplify the representation, we choose the lattice constants as the units of time and length,
resulting in ${\bf e}_t = (1,0)$ and ${\bf e}_x=(0,1)$.

Suppose that the $2$-by-$2$ matrix $A$ represents
a mixing transformation. In this study, we focus on
cyclic transformations, which are the ones that satisfy $A^M = 1$
for some positive integer $M$ (called the order).
By imposing the cyclic condition, we significantly reduce
the number of symmetry groups, enabling us to exhaustively
examine their representations. An arbitrary symmetry transformation
can be expressed as a combination of $A$ and translation.
We denote this combined transformation as $P(j, m, n)$,
which acts on the coordinates as follows:
\be
\begin{split}
\left(\begin{array}{c}t' \\ x' \end{array}\right)= \ P(j,m,n) \  \left(\begin{array}{c}t \\ x \end{array}\right)
=  \ A^j \left(\begin{array}{c}t \\ x \end{array}\right) + 
\left(\begin{array}{c} m \\ n \end{array}\right) ,
\end{split}
\ee
where $j,m$ and $n$ are integers. $P(j,m,n)$ denotes $j$ times of
mixing transformation followed by a translation of
$m$ units in time and $n$ units in space.
A symmetry group is a set of $P$s that meet the group axioms.
The closure under multiplication requires
that the spacetime lattice $\left\{ \left(m, n\right)^T \left| m,n \in \mathbb{Z}
 \right.\right\}$ must keep invariant under $A$.
Together with the existence of inverse element,
we infer that the order of $A$ can only be
$2,3,4$ or $6$~\cite{OurSI}.
%This is reminiscent of the fact that
%there exist only two, three, four and six-fold rotation axis in the crystallography.
The symmetry group can be expressed as
\be
\label{eq:Pexp} 
\mathcal{P} = \left\{ P(j,m,n) \left| j=0, 1, \cdots, M-1; \ \  m,n\in \mathbb{Z}\right. \right\}
\ee
with $M=2,3,4$ or $6$. For given $M$,
the group $\mathcal{P}_M$ is uniquely determined by $A$.
The possible $A$s in $\mathcal{P}_2, \mathcal{P}_3, \mathcal{P}_4$
or $\mathcal{P}_6$ are given in the supplementary materials.

A few examples can help us understand the mixing group.
The form of $A$ in $\mathcal{P}_2$ or $\mathcal{P}_4$ is shown
in Tab.~\ref{tab}, where $a,b,c$ are integers satisfying $bc=-a^2 \pm 1$,
respectively. If $a=0$ and
$b=c=1$, then $A$ is the exchange of $t$ and $x$ (dubbed $A_e$) and
belongs to $\mathcal{P}_2$. If $a=0, b=1$ and $c=-1$,
then $A$ represents a rotation in the $t$-$x$ plane by $90^\circ$ (dubbed
$A_r$) and belongs to $\mathcal{P}_4$.
Figure~\ref{fig:stlattice}a schematically illustrates the operations of $A_e$ and $A_r$.
The transformation $A$ conserves the area of the parallelogram formed
by two noncollinear vectors, because $\text{det} \left(A\right)= \pm 1$. But
$A$ does not necessarily conserve the Euclidean length
of a vector (e.g., consider the case $a=2,b=-3$ and $c=1$).
Therefore, $A$ can be not only rotation, reflection or inversion in the $t$-$x$ plane,
but also nonorthogonal transformations. Notice that $A$
is distinguished from the discrete Lorentz
transformation~\cite{Wang18}, as the latter does not have a finite order.

\emph{Floquet-Bloch band theory.}---
Each quantum theory is a unitary representation of its
corresponding symmetry group. In our case, we aim to
construct the unitary representations of $\mathcal{P}_M$,
and we follow a similar approach as described in
Ref.~[\onlinecite{Wang21}]. To denote the unitary
operator of $P(j,m,n)$, we use $\hat{U}(j,m,n)$,
which follows the same multiplication rule as $P$.
The translation operators $\hat{U}(0,m,n)$ commute
with each other and share common eigenstates.
In the Floquet-Bloch band theory, the eigenstates of
translations are typically represented as $\ket{k,\alpha}$,
where $k \in [ -\pi, \pi)$ denotes the quasi-momentum,
$\alpha$ is the band index, and the corresponding
quasi-energy is denoted as $E_\alpha(k)$ with $E_\alpha(k) \in [ -\pi, \pi)$.
When the operator $\hat{U}(0,m,n)$ acts on $\ket{k,\alpha}$,
it results in $e^{i E_\alpha(k) m - i k n} \ket{k,\alpha}$.
The pair $\left(k,E_\alpha(k)\right)$ represents a point in
the Floquet-Bloch-Brillouin zone (FBBZ), which is topologically
equivalent to a torus (see Fig.~\ref{fig:stlattice}b,c).
The DR of each continuous Floquet-Bloch band,
i.e., the set of $\left(k,E_\alpha(k)\right)$ points, forms a loop on the torus.

Since any element in $\mathcal{P}_M$ can be factorized into
$P(j,m,n)=P(0,m,n)P(j,0,0)$, the representation of
$\mathcal{P}_M$ can be determined by examining the
action of the mixing transformation operator $\hat{U}(j,0,0)$
on the basis states $\ket{k,\alpha}$. Note that
the single-particle Hilbert space is spanned by $\ket{k,\alpha}$.
To determine the representation, it is sufficient to
investigate the action of $\hat{U}(1,0,0)$ since $\hat{U}(j,0,0) = \hat{U}(1,0,0)^j$.
For this purpose, we utilize the multiplication rule:
$P(0,m',n')P(1,0,0) = P(1,0,0) P(0,m,n) $~\cite{OurSI}, or equivalently,
\be
\label{eq:Uex}
\hat{U}(0,m',n') \hat{U}(1,0,0) = \hat{U}(1,0,0) \hat{U}(0,m,n),
\ee
where $(m',n')^T= A \ (m,n)^T$. Acting both sides of Eq.~\eqref{eq:Uex}
on $\ket{k,\alpha}$, we find that $\hat{U}(1,0,0) \ket{k,\alpha}$ is
also an eigenstate of translation operators,
denoted by $\ket{k',\alpha'} =\hat{U}(1,0,0) \ket{k,\alpha} $
without loss of generality. And Eq.~\eqref{eq:Uex} determines a relation
between $k$ and $k'$~\cite{OurSI}, which reads
\be
\label{eq:Aondis}
\left(\begin{array}{c}
k' \\ E_{\alpha'}(k')
\end{array} \right) = \bar{A}
\left(\begin{array}{c}
k \\ E_\alpha(k)
\end{array} \right) \ \ \ \left(\text{mod} \ 2\pi \right),
\ee
where $\bar{A}= \text{det}\left(A\right) A$. Especially,
we find $\bar{A} = -A$ and $\bar{A} = A$ for the symmetry classes $\mathcal{P}_2$
and $\mathcal{P}_4$, respectively. The modulo operation in Eq.~\eqref{eq:Aondis} ensures that $\left(
k', E_{\alpha'}(k')\right)$ falls within the FBBZ. According to definition,
$\bar{A}$ is invertible, and then it is a one-to-one continuous
map between FBBZ and itself. In other words, $\bar{A}$ acts
as a homeomorphism on the FBBZ.

Equation~\eqref{eq:Aondis} reveals that each point
$\left(k,E_\alpha(k)\right)$ within the DRs is mapped by
$\bar{A}$ to another point within the DRs. $\bar{A}$
establishes a one-to-one correspondence between
the set of points within the DRs and itself. In a spacetime crystal
with $N$ continuous bands, each with its corresponding DR
as a loop on the FBBZ torus, $\bar{A}$ acts as
a homeomorphism. Consequently, the image of a loop
(DR) under $\bar{A}$ is guaranteed to be another loop (DR).
Thus, $\bar{A}$ maps each DR loop to another DR loop,
effectively acting as a permutation of the $N$ bands.

%Equation~\eqref{eq:Aondis} tells us, each point $\left(k,E_\alpha(k)\right)$
%in the DRs is mapped by $\bar{A}$ into another point in the DRs. $\bar{A}$ is a one-to-one map
%between the set of points in the DRs and themselves. Suppose a spacetime crystal has
%$N$ continuous bands, whose DRs are $N$ loops on the FBBZ torus.
%$\bar{A}$ is a homeomorphism, therefore,
%the image of a loop (DR) under $\bar{A}$ must be
%another loop (DR). $\bar{A}$ maps each DR loop into another DR loop, therefore, it is
%also a permutation of $N$ bands.

\begin{figure}[htp]
\centering
\vspace{0.2cm}
\includegraphics[width=.45\textwidth]{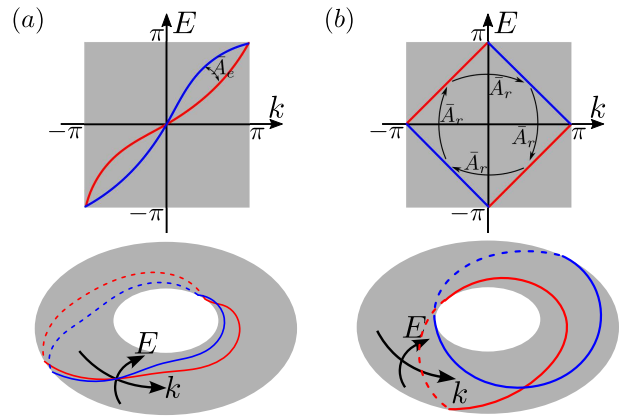}
\caption{The blue and red lines represent a doublet,
consisting of a pair of DRs that are mutually mapped by
(a) $\bar{A}_e$ (exchange transformation) and (b)
$\bar{A}_r$ (rotation transformation). The bottom panels
schematically depict the topological equivalence of these doublets on the FBBZ torus.}
\label{fig:topoDR}
\end{figure}

\emph{Topology of dispersion relation.}---
Equation~\eqref{eq:Aondis} is the sufficient and necessary condition
for a unitary representation of $\mathcal{P}_M$.
Constructing a representation
involves in finding the $\bar{A}$-invariant DRs (solution of Eq.~\eqref{eq:Aondis}).
For general $\bar{A}$, these DRs can be highly nontrivial.
For instance, if $A$ is the exchange $A_e$,
then $\bar{A}_e$ exchanges the quasi-momentum
and quasi-energy. Our familiar DRs, such as quadratic or trigonometric functions,
are not $\bar{A}$-invariant. The nontriviality of $\bar{A}$-invariant
DRs arises from the fact that their loops exhibit nontrivial topology.
The topology of a loop on a
torus is characterized by a pair of integers, which corresponds
to the fundamental group of the torus. As the DR is a continuous
function of $k$ within the range of $[-\pi,\pi)$, a DR loop
must wind around the torus exactly once in the $k$-direction.
The topology of a DR loop is denoted as $\left(1, w\right)$,
where $w$ represents the number of times the DR winds
around the torus in the positive $E$-direction while
completing one revolution in the positive $k$-direction.
It is important to highlight that $w$ has long been recognized
as the average particle displacement over one period.
In each cycle, $w$ units of charge are pumped through the system~\cite{Kitagawa10}.

If $\bar{A}$ maps band $\alpha$ to $\alpha'$, their DRs'
winding numbers ($w_\alpha$ and $w_{\alpha'}$) are
connected to each other according to~\cite{OurSI}
\be
\label{eq:windA}
\pm \left(\begin{array}{c} 1 \\ w_{\alpha'} \end{array}\right) = 
\bar{A} \left(\begin{array}{c} 1 \\ w_{\alpha} \end{array}\right).
\ee
Equation~\eqref{eq:windA} is our key result, which constrains the topology of an
$\bar{A}$-invariant DR. It has no solution for $\bar{A}$ in
$\mathcal{P}_3$ or $\mathcal{P}_6$~\cite{OurSI}, indicating that
these symmetry classes have no representation with continuous bands.
By substituting the expression of $A$ into Eq.~\eqref{eq:windA},
we obtain the band classification for $\mathcal{P}_2$ and $\mathcal{P}_4$.
In $\mathcal{P}_2$, bands are classified as singlets
or doublets. A singlet remains invariant under $\bar{A}$,
while a doublet consists of two bands mapped by $\bar{A}$ into each other,
sharing the same winding number. For $\mathcal{P}_4$,
bands are classified as doublets with odd-function DRs of $k$
and quadruplets (quartets of four bands).
%One band and its image under $\bar{A}^2=-1$ have the same winding.
There are no singlet bands since $w_{\alpha'}=w_\alpha$ constradicts
Eq.~\eqref{eq:windA}. Additionally, the two
bands in a doublet have different winding numbers.
Table~\ref{tab} summarizes the classification of Floquet-Bloch bands with mixing symmetry.

Except for $A$ with $a=\pm1$ (unconventional
space inversion or time reversal), the DR's winding number
must be nonzero~\cite{OurSI}. Figure~\ref{fig:topoDR} presents
examples of DRs that satisfy Eq.~\eqref{eq:windA}.
Figure~\ref{fig:topoDR}(a) shows a doublet pair of bands,
mapped into each other by $\bar{A}_e$ in $\mathcal{P}_2$.
Both bands have a winding number of $+1$. Figure~\ref{fig:topoDR}(b)
shows a doublet pair of bands, mapped into
each other by $\bar{A}_r$ in $\mathcal{P}_4$.
The two bands have winding numbers of $+1$ and $-1$, respectively.

The simplest DRs with mixing symmetries (solution of
Eq.~\eqref{eq:Aondis}) are linear ones: $E(k)=wk$,
where $w= \pm 1, \pm 2,\cdots$ represents the winding number.
From Eqs.~\eqref{eq:Aondis} and~\eqref{eq:windA}, we can fully determine
the mixing symmetries of any linear Floquet-Bloch band~\cite{OurSI}.
In the $\mathcal{P}_2$ symmetry class, a linear band is a singlet,
which keeps invariant under the map $\bar{A}$ with elements
satisfying $a=-b w\pm 1$ and $c=-bw^2 \pm 2w$.
On the other hand, in the $\mathcal{P}_4$ symmetry class,
two bands $E(k)=wk$ and $E'(k)=w'k$ can form a doublet, where $w'=w \pm 1$
or $w'=w\pm 2$~\cite{OurSI}.

By utilizing the $\bar{A}$-invariant $E_\alpha(k)$,
we can readily establish the many-body quantum
theory by introducing the creation operator
$\hat{c}^\dag_{k,\alpha}$ and the annihilation operator
$\hat{c}_{k,\alpha}$ and expressing
the symmetry operators in terms of them~\cite{Wang21}. Specifically, the time translation
operator is $\hat{U}(0,1,0)=e^{i \hat{H}_F}$, where $\hat{H}_F$ is
the effective Floquet Hamiltonian:
\be
\label{eq:HF}
\hat{H}_F = \sum_{k,\alpha} E_\alpha(k) \ \hat{c}^\dag_{k,\alpha}
\hat{c}_{k,\alpha}.
\ee
Note that the symmetry condition of DRs is independent of
whether the particles are bosons or fermions. The operators $\hat{c}_{k,\alpha},\hat{c}_{k,\alpha}^\dag$
are either commutative or anti-commutative, depending on the
species of particles.

A few comments are necessary. First, we ignore the interaction between
particles in the model~\eqref{eq:HF}. Constructing an interacting
theory is significantly more challenging and falls beyond
the scope of the current paper, as the mixing symmetry
imposes constraints not only on the DR
but also on the particle interactions. Second,
for an exhaustive enumeration of quantum theories,
we should also consider the possibility
of $\hat{U}(1,0,0)$ being an anti-unitary operator,
which is discussed in the supplementary materials.

\emph{Realization with local $\hat{H}(t)$.}---
We aim at realize a given Floquet Hamiltonian $\hat{H}_F$,
or equivalently the energy band $E(k)$,
in a cold atom system on an optical lattice.
To achieve experimental feasibility, it is crucial that the time-periodic Hamiltonian
$\hat{H}(t)$ possesses locality in real spacetime.
However, this poses a challenge due to the nonzero winding numbers of the DRs,
which is a characteristic feature of mixing symmetry.

Let us consider a specific example: the linear DRs $E(k)=wk$.
Upon Fourier transformation, the Floquet Hamiltonian
$\hat{H}_F$ contains infinitely-long-range hopping terms in real space, making
them currently inaccessible using existing technology.
In fact, if $\hat{H}(t)$ exhibits locality (i.e., short-range hopping) and simultaneously
maintains space translational symmetry at each time $t$, then the DRs of
$\hat{H}_F$ must possess zero winding~\cite{OurSI}. Therefore,
in order to have an $\bar{A}$-invariant DR, $\hat{H}(t)$ must break
instantaneous translational symmetry.

To design $\hat{H}(t)$ for a given DR, we employ
the recently developed QQFT protocol~\cite{Wang22},
which gives rise to highly flexible Hamiltonian engineering so that
the DRs become completely programmable and the
long-range tunnelings in $\hat{H}_F$ become accessible to
optical lattice experiments. In one period,
denoted as $[0,1)$, the time-dependent Hamiltonian is expressed as
\be
\label{eq:Hseq}
\hat{H}(t)= \sum_{p=1}^D I_p(t) \ \hat{H}_{p} .
\ee
Here, $I_p(t)$ is the indicator function, which is defined as $1$
for $t\in [\left(p-1\right)/D,p/D)$ and zero elsewhere. The parameter $D$ represents the depth
of the Hamiltonian sequence. Each $\hat{H}_p$ contains only
onsite potentials and nearest-neighbor hoppings.
It can generally be written as $\hat{H}_p = \sum_{x}  \left(g^{(p)}_{x} 
\hat{\psi}^\dag_{x} \hat{\psi}_{x+1}  + 
u^{(p)}_{x} \hat{\psi}^\dag_{x} \hat{\psi}_{x} + h.c.\right) $,
where $\hat{\psi}^\dag_{x}$ and $\hat{\psi}_{x}$ are the creation
and annihilation operators at site $x$, respectively, and $g$ and $u$ denote the hopping strength
and onsite potentials, respectively.
The unitary evolution over one period
is given by $e^{-i\hat{H}_F} = e^{- i \hat{H}_D/D} \cdots e^{- i \hat{H}_2/D}
e^{- i \hat{H}_1/D}$. For a lattice model with $L$ sites,
the depth $D$ scales as $ L \log L$ for large $L$~\cite{Wang22}, in the QQFT protocol.
As the system size increases, the effort required for simulation grows super-linearly.
In recent developments in cold atom technology,
spatially resolved control of the atom-confining potential has been
achieved, enabling the realization of
a sequence of local Hamiltonians like Eq.~\eqref{eq:Hseq}.
It has been shown that systems with sizes
up to several tens of sites are accessible in present experiments~\cite{Qiu20,Wang22}.

\begin{figure}[htp]
\centering
\vspace{0.2cm}
\includegraphics[width=.45\textwidth]{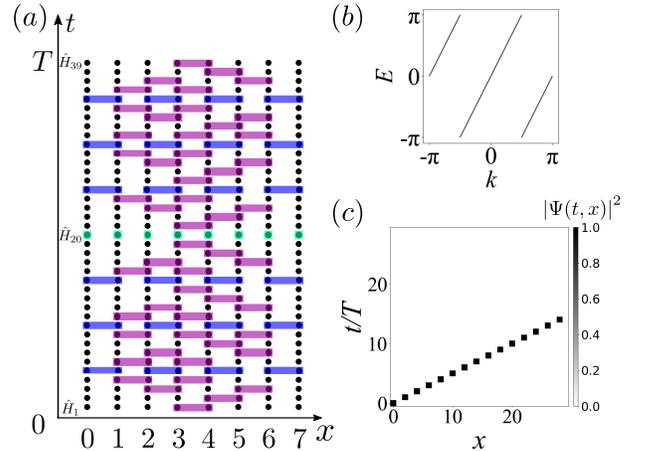}
\caption{(a) The sequence of $\hat{H}_p$ operations
with $p=1,\cdots, 39$ in one period ($t\in \left[0,T\right]$)
for a lattice of size $L=8$. The lattice sites,
labeled as $0\sim 7$, are represented by black dots.
The purple and blue rectangles depict the swap and local Fourier
operations between neighboring sites, respectively.
The green squares represent the onsite-potential operation.
(b) The linear dispersion relation $E=2k$. (c)
The probability density $\left|\Psi(t,x)\right|^2$ as a particle propagates.}
\label{fig:prop}
\end{figure}

Figure~\ref{fig:prop}(a) illustrates the sequence
of $\hat{H}_p$ operations that generate $E(k)=wk$ on a chain of length $L=8$.
Within each period, a total of $39$ operations are performed, including $32$ swaps between
neighboring sites, $6$ local Fourier transformations, and one evolution of
the onsite potential. For more detailed information, please refer to the supplemental material.

\emph{Mixing symmetry in the wave function.}---
To observe the mixing symmetry, one can utilize the fact that
the mixing symmetry manifests itself in the particle propagator in real spacetime.
For a particle initially located at position $x=0$ and time $t=0$, its
wave function at a later time (multiples of the period) satisfies:
\be\label{eq:psialphatx}
\Psi_\alpha(t,x) = \Psi_{\alpha'} \left( t',x'\right),
\ee
where $\left(t',x'\right)^T = A\left(t,x\right)^T$ and $t,x$ are arbitrary integers.
$\alpha'$ is the map of band-$\alpha$ under $\bar{A}$.
For $\alpha'=\alpha$ (a singlet band in the $\mathcal{P}_2$ class),
Eq.~\eqref{eq:psialphatx} imposes a strong constraint on the wave function.
For $\alpha' \neq \alpha$, Eq.~\eqref{eq:psialphatx} provides a
connection between the wave functions in different bands.

For a concrete example, let us see a linear band with $E(k)=wk$, in which
the particle moves at a constant speed, just like a classical particle.
The wave function is calculated to be $\Psi_\alpha(t,x)=\delta_{x,w t}$.
In previous discussions, we already show that such a band exhibits the
$\mathcal{P}_2$ symmetry when the elements of $A$ are
$a = -b w\pm 1$ and $c=-bw^2 \pm 2 w$. It is easy to verify that
$\delta_{x,w t}$ does remain invariant as $\left(t,x\right)^T$ transforms
under $A$. Using the QQFT protocol, we perfectly repeat the evolution of
wave function on a lattice of length $L=2^l$.
Figure~\ref{fig:prop}(b) displays the DR as $w=2$, and Fig.~\ref{fig:prop}(c)
displays the corresponding wave function in the QQFT simulation.

The probability distribution of particles, i.e. $\left|\Psi(t,x)\right|^2$, obviously
meets the same symmetry as shown in Eq.~\eqref{eq:psialphatx}.
In experiments, instead of a single particle,
one can use the Bose-Einstein condensate (BEC) for
observation, and then, $\left| \Psi(t,x)\right|^2$ represents the density of atoms.
The density distribution forms a symmetric pattern which remains
invariant under $A$, which will be a smoking gun
signal of mixing symmetry.

\emph{Discussion.}---
This paper presents an innovative discovery of Floquet-Bloch
band theories that exhibit a unique mixing symmetry, which
intertwines the space and time coordinates. We provide a comprehensive
classification of Floquet-Bloch bands based on the cyclic mixing
transformations of finite order. Notably, only the groups
$\mathcal{P}_2$ and $\mathcal{P}_4$ possess continuous
representations, where the mixing symmetry imposes constraints
on the dispersion relation of each band.
Furthermore, we reveal that the winding number of the dispersion
relation on the Floquet-Bloch-Brillouin torus must adhere to
a symmetry condition. To achieve a non-zero winding number,
it is essential for the time-dependent Hamiltonian of the theory
to break the instantaneous translation symmetry, a feat attainable
through the implementation of QQFT on an optical lattice.
Remarkably, the mixing symmetry manifests in the atom density,
which becomes experimentally measurable, demonstrating its
impact on the spacetime distribution.
This discovery unveils a broader symmetry family that has been
previously ignored, as the mixing symmetry transcends
pure spatial or temporal characteristics and instead establishes
correlations between space and time. Its exploration enhances
our comprehension of symmetry in crystals. Looking ahead,
intriguing open questions include the investigation of noncyclic
mixing symmetry and the exploration of mixing-symmetry protected topological states of matter.

\begin{acknowledgments}
{\it Acknowledgement.---} 
The work is supported by National Natural Science Foundation of
China (Grants Nos. 11835011, 11774315),
and the Junior Associates program of the Abdus Salam International Center for Theoretical Physics. 
We thank X. Wang for useful discussions.
\end{acknowledgments}

%\bibliography{references}

\begin{thebibliography}{35}
\expandafter\ifx\csname natexlab\endcsname\relax\def\natexlab#1{#1}\fi
\expandafter\ifx\csname bibnamefont\endcsname\relax
  \def\bibnamefont#1{#1}\fi
\expandafter\ifx\csname bibfnamefont\endcsname\relax
  \def\bibfnamefont#1{#1}\fi
\expandafter\ifx\csname citenamefont\endcsname\relax
  \def\citenamefont#1{#1}\fi
\expandafter\ifx\csname url\endcsname\relax
  \def\url#1{\texttt{#1}}\fi
\expandafter\ifx\csname urlprefix\endcsname\relax\def\urlprefix{URL }\fi
\providecommand{\bibinfo}[2]{#2}
\providecommand{\eprint}[2][]{\url{#2}}

\bibitem{Oka09} T. Oka and H. Aoki, Phys. Rev. B {\bf 79}, 081406(R) (2009).
\bibitem{Kitagawa10} T. Kitagawa, E. Berg, M. Rudner, and E. Demler,
Phys. Rev. B {\bf 82}, 235114 (2010).
\bibitem{Lindner11} N. H. Lindner, G. Refael, and V. Galitski, Nat. Phys. {\bf 7}, 490 (2011).
\bibitem{Cooper19} N. R. Cooper, J. Dalibard, and I. B. Spielman, Rev. Mod. Phys. {\bf 91}, 015005 (2019).
\bibitem{Rudner20} M. S. Rudner and N. H. Lindner, Nat. Rev. Phys. {\bf 2}, 229 (2020).
\bibitem{Rechtsman13} M. C. Rechtsman, J. M. Zeuner, Y. Plotnik, Y. Lumer, S. Nolte,
M. Segev, and A. Szameit, Nature {\bf 496}, 196 (2013).
\bibitem{Wang13} Y. H. Wang, H. Steinberg, P. Jarillo-Herrero, and N. Gedik,
Science {\bf 342}, 453 (2013).
\bibitem{Jotzu14} G. Jotzu, M. Messer, R. Desbuquois, M. Lebrat, T. Uehlinger,
D. Greif, and T. Esslinger, Nature {\bf 515}, 237 (2014).
\bibitem{Sorensen05} A. S. S{\o}rensen, E. Demler, and M. D. Lukin. Phys. Rev. Lett. {\bf 94}, 086803 (2005).
\bibitem{Eckardt05} A. Eckardt, C. Weiss, and M. Holthaus, Phys. Rev. Lett. {\bf 95}, 260404 (2005).
\bibitem{Goldman14} N. Goldman and J. Dalibard, Phys. Rev. X {\bf 4}, 031027 (2014).
\bibitem{Oka19} T. Oka and S. Kitamura, Annu. Rev. Condens. Matter Phys. {\bf 10}, 387 (2019).
\bibitem{Else16} D. V. Else, B. Bauer, and C. Nayak,
Phys. Rev. Lett. {\bf 117}, 090402 (2016).
\bibitem{Khemani16} V. Khemani, A. Lazarides, R. Moessner, and S. L. Sondhi,
Phys. Rev. Lett. {\bf 116}, 250401 (2016).
\bibitem{Yao17} N. Y. Yao, A. C. Potter, I.-D. Potirniche, and A. Vishwanath,
Phys. Rev. Lett. {\bf 118}, 030401 (2017).
\bibitem{Ponte15} P. Ponte, Z. Papi\'{c},
F. Huveneers, and D. A. Abanin, Phys. Rev. Lett. {\bf 114}, 140401 (2015).
\bibitem{Lazarides15} A. Lazarides, A. Das, and R. Moessner, Phys. Rev. Lett. {\bf 115}, 030402 (2015).
\bibitem{Bordia17} P. Bordia, H. L\"{u}schen, U. Schneider, M. Knap, and I. Bloch,
Nat. Phys. {\bf 13}, 460 (2017).
\bibitem{Rudner13} M. S. Rudner, N. H. Lindner, E. Berg, and M. Levin,
Phys. Rev. X {\bf 3}, 031005 (2013).

\bibitem{Ashcroft76} N. W. Ashcroft and N. D. Mermin, {\it Solid state physics}
(Tomson Learning Inc., London, UK, 1976).
\bibitem{Altland97} A. Altland and M. R. Zirnbauer,
Phys. Rev. B {\bf 55}, 1142 (1997).
\bibitem{Schnyder08} A. P. Schnyder, S. Ryu, A. Furusaki, and A. W. W. Ludwig,
Phys. Rev. B {\bf 78}, 195125 (2008).
\bibitem{Kitaev09} A. Kitaev, AIP Conf. Proc. {\bf 1134}, 22 (2009).
\bibitem{Nathan15} F. Nathan and M. S Rudner, New J. Phys. {\bf 17}, 125014 (2015).
\bibitem{Potter16} A. C. Potter, T. Morimoto, and A. Vishwanath,
Phys. Rev. X {\bf 6}, 041001 (2016).
\bibitem{Else16b} D. V. Else and C. Nayak, Phys. Rev. B {\bf 93}, 201103(R) (2016).
\bibitem{Roy17} R. Roy and F. Harper, Phys. Rev. B {\bf 96}, 155118 (2017).
%Topological quantum chemistry
\bibitem{Bradlyn17} B. Bradlyn, L. Elcoro, J. Cano, M. G. Vergniory,
Z. Wang, C. Felser, M. I. Aroyo, and B. A. Bernevig, Nature {\bf 547}, 298 (2017).
\bibitem{Po17} H. C. Po, A. Vishwanath, and H. Watanabe, Nat. Commun.
{\bf 8}, 50 (2017).
\bibitem{Kruthoff17} J. Kruthoff, J. de Boer, J. van Wezel, C. L. Kane, and R.-J. Slager,
Phys. Rev. X {\bf 7}, 041069 (2017).

\bibitem{Morimoto17} T. Morimoto, H. C. Po, and A. Vishwanath, Phys. Rev. B {\bf 95}, 195155 (2017).
\bibitem{Xu18} S. Xu and C. Wu, Phys. Rev. Lett. {\bf 120}, 096401 (2018).
\bibitem{Peng19} Y. Peng and G. Refael, Phys. Rev. Lett. {\bf 123}, 016806 (2019).
\bibitem{Mochizuki20} K. Mochizuki, T. Bessho, M. Sato, and H. Obuse,
Phys. Rev. B {\bf 102}, 035418 (2020).

\bibitem{Wang18} P. Wang, New J. Phys. {\bf 20}, 023042 (2018).
\bibitem{Wang20} X. Li, J. Chai, H. Zhu, and P. Wang, J. Phys.: Condens. Matter {\bf 32},
145402 (2020).
\bibitem{Wang21} P. Wang, J. Phys. A: Math. Theor. {\bf 54}, 115003 (2021).
\bibitem{Wang22} P. Wang, Z. Huang, X. Qiu, and X. Li, Phys. Rev. B {\bf 106}, 134313 (2022).

\bibitem{2016_Weiss_Science} Y. Wang, A. Kumar, T.-Y. Wu, and D. S. Weiss,
Science {\bf 352}, 1562 (2016).
\bibitem{browaeys2020many} A. Browaeys and T. Lahaye, Nat. Phys. {\bf 16}, 132 (2020).

\bibitem{OurSI} See Supplementary Materials.

\bibitem{Weinberg} S. Weinberg, {\it The Quantum Theory of Fields} (Cambridge
University Press, Cambridge, England, 1995).

\bibitem{Qiu20} X. Qiu, J. Zou, X. Qi, and X. Li, npj Quantum Inf. {\bf 6}, 87 (2020).

%\bibitem{Thouless83} D. J. Thouless, Phys. Rev. B {\bf 27}, 6083 (1983).

\end{thebibliography}

\clearpage

\begin{widetext} 
%\newpage 
\renewcommand{\theequation}{S\arabic{equation}}
\renewcommand{\thesection}{S-\arabic{section}}
\renewcommand{\thefigure}{S\arabic{figure}}
\renewcommand{\thetable}{S\arabic{table}}
\setcounter{equation}{0}
\setcounter{figure}{0}
\setcounter{table}{0}

\newpage

\begin{center} 
{\Huge \bf Supplementary Materials} \\
\end{center} 

\section{Mixing groups}

According to definition, the mixing symmetry group has two important subgroups. One is
the cyclic group that contains the mixing transformations, i.e.,
$\mathcal{A}=\left\{1,A,A^2,\cdots, A^{M-1} \right\}$ with $M$ being the order.
The other consists of the translations, reading $\mathcal{T}= \left\{ 
\left(m,n\right) \left| m,n \in \mathbb{Z} \right.\right\}$. Usually, the group
that has $\mathcal{A}$ and $\mathcal{T}$ as subgroups is not unique.
In this paper, we only consider the symmorphic group, which is the direct
product of $\mathcal{A}$ and $\mathcal{T}$. The group element is written
as $P(j,m,n)$, which represents the mixing transformation $A^j$ followed by the translation
of vector $(m,n)$. It is easy to see that, $\mathcal{P}=\left\{ P(j,m,n)\right\}$
is a group if and only if the spacetime lattice $\mathcal{T}$ keeps invariant under $A$.
Because $A$ is invertible ($A^{-1}=A^{M-1}$), $\mathcal{T}$ keeps invariant under $A$
if and only if $m'$ and $n'$, defined by $\left(m',n'\right)^T = A\left(m,n\right)^T $, are integers
for arbitrary $m,n\in \mathbb{Z}$. Furthermore, this condition can be simplified into
$A\left(1,0\right)^T $ and $A\left(0,1\right)^T $ being integer pairs.

We generally express the matrix $A$ and its inverse as
\be
\label{eq:app:aainv}
A=\left(\begin{array}{cc} a_{11} & a_{12} \\ a_{21} & a_{22} \end{array}\right) \ \ \ \text{and}
\ \ \  A^{-1} = \frac{1}{\text{det} \left(A\right)}
\left(\begin{array}{cc} a_{22} & -a_{12} \\ -a_{21} & a_{11} \end{array}\right),
\ee
respectively. Then, the condition that $A\left(1,0\right)^T $ and
$A\left(0,1\right)^T $ are integer pairs translates into
$a_{11}, a_{12}, a_{21}, a_{22}$ being all integers. But $A^j \left(1,0\right)^T $ and
$A^j \left(1,0\right)^T $ must be also integer pairs for $j=2,3,\cdots, M-1$.
The case of $j=M-1$, or equivalently $j=-1$, is especially important, from which we derive
that $a_{11}/\text{det} \left(A\right), a_{12}/\text{det} \left(A\right), 
a_{21}/\text{det} \left(A\right), a_{22}/\text{det} \left(A\right)$ are integers.
For $a_{ij}$ and $a_{ij}/\text{det} \left(A\right)$ to be both integers, we require
$\text{det} \left(A\right) = \pm 1$. To see it, one can use proof by contradiction
(the assumption $\text{det} \left(A\right)=\pm 2, \pm 3, \cdots$ leads to contradiction).

To find all the cyclic $A$s, we study the
eigenvalues of $A$, i.e. a pair of complex numbers expressed as $\lambda_\pm
= \frac{a_{11}+a_{22}}{2 }\pm \sqrt{\left(\frac{a_{11}+a_{22}}{2 }\right)^2- \text{det}\left(A\right)}$.
The cyclic condition ($A^M=1$) indicates $\left| \lambda_\pm \right| \equiv 1$,
which is possible only if $a_{11}+a_{22} = 0, \pm 1 ,\pm 2$.
As $a_{11}+a_{22} = 0$ and $\text{det}\left(A\right)=-1$, a straightforward
calculation shows $A^2=1$. Such $A$s can be written in a more compact form as
\be
\label{eq:app:A24}
A=\left(\begin{array}{cc} a & b \\ c & -a \end{array}\right),
\ee
where $a,b,c$ are arbitrary integers satisfying
$bc = -a^2+1$. As $a_{11}+a_{22} = 0$ and $\text{det}\left(A\right)=+1$,
we find $A^4=1$, and $A$ has the same expression as Eq.~\eqref{eq:app:A24}
but with $bc = -a^2-1$. As $a_{11}+a_{22} = \pm 1$, only $\text{det}\left(A\right)=1$
is consistent with $\left| \lambda_\pm \right| \equiv 1$ but $\text{det}\left(A\right)=-1$
is not, and we find $A^6=1$ or $A^3=1$, respectively. As $a_{11}+a_{22} = \pm 2$,
the calculation shows that there does not exist a finite $M$ so that $A^M=1$,
except for $A=\pm 1$, which is trivial and then ignored.

To summarize, the values of $M$ are $2,3,4$ or $6$, and the
corresponding symmetry groups are denoted by $\mathcal{P}_2,\mathcal{P}_3,
\mathcal{P}_4$ or $\mathcal{P}_6$, respectively. For a given $M$, $\mathcal{P}_M$
is a class of groups, with different groups having different $A$.
In $\mathcal{P}_2$, $A$ is the matrix~\eqref{eq:app:A24} with $bc=-a^2+1$.
In $\mathcal{P}_4$, $A$ is the matrix~\eqref{eq:app:A24} with $bc=-a^2-1$.
In $\mathcal{P}_3$, $A$ is the matrix~\eqref{eq:app:aainv} with
the components being arbitrary integers that satisfy
$a_{11}+a_{22} = - 1$ and $a_{11}a_{22}-a_{12}a_{21}=1$.
In $\mathcal{P}_6$, $A$ is the matrix~\eqref{eq:app:aainv} with
the components being arbitrary integers that satisfy
$a_{11}+a_{22} = + 1$ and $a_{11}a_{22}-a_{12}a_{21}=1$.
 
 \section{Unitary and anti-unitary representations}
 
 We use $\ket{k,\alpha}$ to denote the single-particle
 eigenstate of the translation operators $\hat{U}(0,m,n)$
 with $m,n\in \mathbb{Z}$. According to the Floquet-Bloch band theory,
 without loss of generality, the corresponding eigenvalue can be expressed as
 $e^{imE_\alpha(k)-ikn}$, where $k$ and $E_\alpha(k)$ are the quasi-momentum
 and quasi-energy, respectively, and $\alpha$ is the band index.
 Let us calculate $\hat{U}(1,0,0)\ket{k,\alpha}$. From the definition of $P(j,m,n)$, it is easy
 see $P(0,m',n')P(1,0,0) = P(1,0,0)P(0,m,n)$ with $\left(m',n'\right)^T=
 A\left(m,n\right)^T$. $\hat{U}(j,m,n)$ is the representation of $P(j,m,n)$, then
 they satisfy the same multiplication rule. We obtain
 \be
 \label{eq:app:umpnpu1}
 \hat{U}(0,m',n')\hat{U}(1,0,0) \ket{k,\alpha}= \hat{U}(1,0,0)\hat{U}(0,m,n)
 \ket{k,\alpha} = e^{imE_\alpha(k)-ikn}\hat{U}(1,0,0) \ket{k,\alpha}.
 \ee
 Equation~\eqref{eq:app:umpnpu1} tells us that $\hat{U}(1,0,0) \ket{k,\alpha}$
 is the eigenstate of $ \hat{U}(0,m',n')$ with the eigenvalue being $e^{imE_\alpha(k)-ikn}$.
 But $m'$ and $n'$ can be arbitrary integers, because $\left(m',n'\right)^T=
 A\left(m,n\right)^T$ and $A$ is invertible. $\hat{U}(1,0,0) \ket{k,\alpha}$ is then
 the common eigenstate of the translation operators, denoted by
 $\ket{k',\alpha'}$ without loss of generality. Using the notations $k'$ and $\alpha'$,
 we calculate the left-hand side of Eq.~\eqref{eq:app:umpnpu1} and then obtain
 \be
 \label{eq:app:ekpk}
 \displaystyle e^{im'E_{\alpha '}(k')-ik'n'} = e^{imE_\alpha(k)-ikn}.
 \ee
 Using the fact that $\text{det}\left(A\right)=\pm 1$ and the expression
 of $A^{-1}$ in Eq.~\eqref{eq:app:aainv}, we quickly find
 \be
 \label{eq:app:kpkrela}
\left(\begin{array}{c}
k' \\ E_{\alpha'}(k')
\end{array} \right) = \bar{A}
\left(\begin{array}{c}
k \\ E_\alpha(k)
\end{array} \right) \ \ \ \left(\text{mod} \ 2\pi \right)
\ee
with $\bar{A}= \text{det}\left(A\right) \cdot A = \pm A$. 

In the above derivation, we assume that $\hat{U}(1,0,0)$, i.e. the
representation of $A$, is a unitary operator. To make our discussion
complete, we also need to consider the possibility of $\hat{U}(1,0,0)$
being an anti-unitary operator. In this case, the multiplication rule
keeps the same, but Eq.~\eqref{eq:app:umpnpu1} changes into
$ \hat{U}(0,m',n')\hat{U}(1,0,0) \ket{k,\alpha}=
 e^{-imE_\alpha(k)+ikn}\hat{U}(1,0,0) \ket{k,\alpha}$. Due to the reason
 mentioned above, we still assume $\hat{U}(1,0,0) \ket{k,\alpha}=\ket{k',\alpha'}$.
Then Eq.~\eqref{eq:app:ekpk} becomes $e^{im'E_{\alpha '}(k')-ik'n'} 
= e^{-imE_\alpha(k)+ikn}$. Equation~\eqref{eq:app:kpkrela} keeps
the same but with $\bar{A}= - \text{det}\left(A\right) \cdot A$.
Comparing the anti-unitary representation with the unitary representation,
we find that the dispersion relation satisfies the same equation with only
the sign of $\bar{A}$ changing. On the other hand, if we do the change $A\to -A$
in the unitary representation, the sign of $\bar{A}$
also changes, since $\text{det}\left(A\right)=
\text{det}\left(-A\right)$. Moreover, if $A$ is a cyclic matrix, so is $-A$.
Therefore, for each anti-unitary representation, there exists a unitary
representation that has exactly the same $\bar{A}$, and then the
dispersion relation, i.e. the solution of Eq.~\eqref{eq:app:kpkrela}, is also the same.
The consideration of anti-unitary representation leads to
nothing new in the dispersion relation.

\section{Topology of dispersion relation}

We assume that the Floquet-Bloch band is continuous, or in other words,
$E_\alpha(k)$ is a continuous function of $k$ everywhere in
the Floquet-Bloch-Brillouin zone (FBBZ). The dispersion relation (DR)
of each band is then a loop on the FBBZ torus. The transformation $\bar{A}$
($\text{mod} \ 2\pi$) defined by Eq.~\eqref{eq:app:kpkrela}
maps a point in the FBBZ to another point in the FBBZ.
Furthermore, it is a one-to-one map. Otherwise,
suppose $\left(k_1,E_1\right)\neq \left(k_2,E_2\right)$ are mapped
into the same $\left(k',E'\right)$, then we have $\bar{A} \left(k_1-k_2,E_1-E_2\right)^T
= 2\pi \left( m,n \right)^T$ with $m,n$ being some integers. But
the matrix $\bar{A}=\pm A$ or its inverse always map an integer pair
into another integer pair, and then $\left(k_1-k_2,E_1-E_2\right)
= 2\pi \left( m',n' \right)$ with $m',n'$ being integers. This is impossible
except for $k_1 = k_2$ and $E_1=E_2$, because $\left(k_1,E_1\right)$
and $ \left(k_2,E_2\right)$ are both in the FBBZ.

The one-to-one map $\bar{A}$ is by definition continuous, so is its inverse. $\bar{A}$ is then a
homeomorphism. As a consequence, an arbitrary loop on the FBBZ torus
must be mapped by $\bar{A}$ into another loop.
In the main text, we show that the
spacetime crystal has the mixing symmetry if and only if the single-particle
DRs are $\bar{A}$-invariant. And if the DRs are $\bar{A}$-invariant, then
the DR of a band $\alpha$, i.e. a loop, must
be mapped into the DR of another band $\alpha'$
(it is possible that $\alpha=\alpha'$). 
Note that, from the pure mathematical point of view, it is also possible
that the image of a DR loop is a non-DR loop (e.g., a loop on which $k$
keeps a constant but $E$ travels around the torus once). But in that case,
the DRs are not $\bar{A}$-invariant, and then the corresponding spacetime crystal
has no mixing symmetry, which is uninteresting to us.

Next, we study the topologies of
the DR loops of $\alpha$ and $\alpha'$. Using the knowledge of
the fundamental group of torus, we describe
the topology of a loop by two integers, which are the numbers
of times the loop winds around the torus in the positive $k$- and $E$-directions,
respectively. A DR loop winds around the torus once and only once in the $k$-direction,
otherwise, there would exist some $k\in [-\pi,\pi)$ at which $E(k)$ has no definition or
has multiple values, which contradicts with the fact that $E(k)$ is a function of $k$ defined
in the domain $[-\pi,\pi)$. Therefore, the topology of the $\alpha$-band DR is
given by the pair $\left(1,w_\alpha\right)$, in which $w_\alpha$ is the number of
times the loop winds in the positive-$E$ direction as it winds once in the positive-$k$
direction.

An easy way of calculating $w_\alpha$ is by depicting $E_\alpha(k)$
in the extended quasi-energy zone, in which the range of $E$
is extended to $\left(-\infty,\infty\right)$ instead of being limited in $[-\pi,\pi)$.
In the extended-zone scheme, we can force $E_\alpha(k)$ to
be continuous in the absence of the modulo operation,
$E_\alpha(k)$ then becomes a curve
in the $k$-$E$ plane with $k\in [-\pi,\pi)$ and $E\in \left(-\infty,\infty\right)$.
The continuity of $E_\alpha(k) \ \left(\text{mod} \ 2\pi\right)$ requires
$\left(E_\alpha(\pi) - E_\alpha(-\pi)\right)$ being an integer times of $2\pi$, and this
integer is exactly $w_\alpha$:
\be
E_\alpha(\pi) - E_\alpha(-\pi) = 2\pi w_\alpha.
\ee

Now, we study the image of $\left\{\left(k,E_\alpha(k)\right)\right\}$ under
the matrix $\bar{A}$, in the extended-zone scheme. Without the
modulo operation, $\bar{A}$ is an invertible one-to-one map in
the $k$-$E$ plane, moreover, it is a linear map. Therefore, when $\left(k,E_\alpha(k)\right)$
starts from the left end $\left(-\pi,E_\alpha(-\pi)\right)$, and goes towards
the right end $\left(\pi,E_\alpha(\pi)\right)$, its image $\left(k',E_{\alpha'}(k')\right)$
draws a curve in the plane. The end points of the image curve
are $\left(k'_0,E_{\alpha'}(k'_0)\right)^T=\bar{A} \left(-\pi,E_\alpha(-\pi)\right)^T$
and $\left(k'_1,E_{\alpha'}(k'_1)\right)^T = \bar{A}  \left(\pi,E_\alpha(\pi)\right)^T$, respectively.
Then, the winding number of $\alpha'$ evaluates $w_{\alpha'} =\left(E_{\alpha'}(k'_1)
- E_{\alpha'}(k'_0)\right)/\left( k'_1- k'_0\right)$. 
An important property of the image curve is that $\left| k'_1-k'_0\right|$ must be $2\pi$.
The proof is as follows. First, the range of $k'$ must be
integer times of $2\pi$, because the image is a complete DR loop
(of band $\alpha'$) after the modulo operation. Second, the range
of $k'$ cannot be $2\pi n$ with $n>1$. Otherwise,
as $\left(k,E_\alpha(k)\right)$ travels around the $\alpha$-DR loop once,
$\left(k',E_{\alpha'}(k')\right)$ already travels around the $\alpha'$-DR loop $n$ times,
which contradicts with the fact that $\bar{A} \ \left(\text{mod} \ 2\pi\right)$
is a one-to-one map on the torus. Based on the above arguments, we derive
\be
\label{eq:app:windA}
\pm \left(\begin{array}{c} 1 \\ w_{\alpha'} \end{array}\right) = 
\bar{A} \left(\begin{array}{c} 1 \\ w_{\alpha} \end{array}\right),
\ee
where "$\pm$" corresponds to $k'_1-k'_0 = \pm 2\pi$, respectively.

\section{Mixing symmetries of linear $E(k)$} 

We determine the mixing symmetries of a linear DR, given by $E(k)=wk$ with $w=\pm 1,\pm 2,\cdots$,
by making the following observation.
If the topology condition $\pm \left(1,w'\right)^T=
\bar{A} \left(1,w\right)^T$ is satisfied,
we can multiply both sides by $k$ to obtain $\left(k',E'\right)^T = \bar{A} \left(k,E\right)^T$,
where $k'=\pm k$ and $E'=w' k'$. Therefore, the topology condition is
sufficient for one linear band to be mapped by $\bar{A}$ into
another linear band.

Let us first consider the $\mathcal{P}_2$ symmetry class. Since $w=w'$
under the map $\bar{A}$ (see Tab.~I of the main text),
a linear $E(k)$ is always mapped into itself and remains a singlet band in the $\mathcal{P}_2$ class.
Using the equation $\pm \left(1,w\right)^T=
\bar{A} \left(1,w\right)^T$ and the expression of $\bar{A}$, we immediately
find $a = -b w\pm 1$ and $c=-bw^2 \pm 2 w$. For a given $w$, there exist
an infinite number of mixing matrices (with different $b$) in the $\mathcal{P}_2$ class:
\be
A= \left(\begin{array}{ccc}-bw \pm 1 & & b \\
-bw^2 \pm 2w & & bw\mp 1 \end{array}\right).
\ee
The linear band $E(k)=wk$ always exhibits $\mathcal{P}_2$ symmetries.

Next, we consider the $\mathcal{P}_4$ symmetry class. Since $E(k)=wk$
is an odd function of $k$, the linear band must be one branch of a doublet
(Tab.~I of the main text). Suppose the DR of its paired band is $E'(k')
=w'k'$. Using $\pm \left(1,w'\right)^T=
\bar{A} \left(1,w\right)^T$ and the expression of $\bar{A}$, we find
$a=-bw\pm 1$, $c=-bw^2\pm 2w -2/b$, and $w'=w\mp 2/b$. $c$ and
$w'$ must be integers, therefore, $b$ can only take the values $\pm 1, \pm 2$.
For a given $w$, there exist $8$ mixing matrices in the $\mathcal{P}_4$
symmetry class:
\be
A= \left(\begin{array}{ccc}-w \pm 1 & & 1 \\
-w^2 \pm 2 w - 2 & & w\mp 1 \end{array}\right), \ 
\left(\begin{array}{ccc}w \pm 1 & & -1 \\
w^2 \pm 2 w + 2 & & -w\mp 1 \end{array}\right), \ 
\left(\begin{array}{ccc}-2w \pm 1 & & 2 \\
-2w^2 \pm 2 w - 1 & & 2w\mp 1 \end{array}\right), \ 
\left(\begin{array}{ccc}2w \pm 1 & & -2 \\
2w^2 \pm 2 w +1 & & -2w\mp 1 \end{array}\right).
\ee
The corresponding $w'$ is given by $w'=w\mp 2, w\pm 2,
w\mp 1, w\pm 1$, respectively.

The above analysis exhausts all the mixing symmetries of a linear band.

\section{Construction of $\hat{H}(t)$} 

Our target is to simulate $\hat{H}_F =  \sum_{k} E(k) \ \hat{c}^\dag_{k}
\hat{c}_{k}$ that has mixing symmetry
by a periodic Hamiltonian $\hat{H}(t)$ with locality. First, we will prove that,
if $\hat{H}(t)$ has both locality and space translation symmetry
at each $t$, then the winding number of $E(k)$ is zero.
For simplicity, we consider a lattice model, in which a set of sites
are spatially located at the coordinates $j= 0, \pm 1, \pm 2, \cdots$, respectively.
In condensed matter community, the lattice models are widely employed
in the study of particles moving in a periodic potential, because it is more
difficult to directly deal with the differential operators in the continuous space.
Without loss of generality, we define
\be
\label{eq:app:Ht}
\hat{H}(t) =\sum_{j} \sum_{\Delta j=-R}^R  f(\Delta j,t)
\hat{\psi}^\dag_{j} \hat{\psi}_{j+\Delta j}  ,
\ee
where $f(\Delta j,t)=f^*\left(-\Delta j,t\right)$ is the
hopping strength, and $\hat{\psi}^\dag$ and $\hat{\psi}$ are
the onsite creation and annihilation operators, respectively.
The assumption that $\hat{H}(t)$ has space translation symmetry
at each moment, is hidden in the fact that $f(t)$ is independent of $j$.
The locality of $\hat{H}(t)$ manifests itself as the existence of
a distance cutoff for hopping. The largest distance over which
there are nonzero hopping terms is set to be $R$. After a Fourier transform,
Eq.~\eqref{eq:app:Ht} changes into $\hat{H}(t) =\displaystyle\sum_{k} 
\displaystyle E(k,t)
\hat{\psi}^\dag_{k} \hat{\psi}_{k}$, where $\label{eq:app:fnnpk}
E(k,t) = \sum_{\Delta j=-R}^R f (\Delta j,t)e^{ik\Delta j}$.
$E(k,t)$ is a sum of finite number of terms, with each term
being a trigonometric function of $k$. If we depict these trigonometric functions
on the FBBZ torus, they all have zero winding number, and then
their sum, i.e. $E(k,t)$, must also have zero winding number.
The Floquet Hamiltonian $\hat{H}_F$ can be obtained by integrating $\hat{H}(t)$ over one period.
Because $\hat{H}(t)$ at different $t$ commutes with each other.
Then we obtain $E(k) = \int^T_0 dt \ E(k,t) /T$ where $T=1$ is the period.
Since $E(k,t)$ at each $t$ has zero winding, $E(k)$ must also have
zero winding.

The DRs of a spacetime crystal with mixing symmetry usually
have nonzero winding. And due to the above arguments,
if we ask $\hat{H}(t)$ to be local and we want
to simulate a $\hat{H}_F$ with nonzero-winding DRs, we
need to break the instantaneous translation symmetry in $\hat{H}(t)$.
In previous theoretical
or experimental studies, people often focus on the atom-confining potential
that keeps the instantaneous translation symmetry. This explains
why the mixing symmetry has not been observed accidentally.
The recently developed digital-micromirror-device and sub-wavelength
techniques have realized programmable instantaneous-translation-symmetry-breaking
potentials in the cold atomic gases. This provides the
foundation for experimentally realizing $\hat{H}(t)$.

Because we already know the DRs of $\hat{H}_F$,
the quadratic quantum Fourier transform (QQFT) protocol is
especially useful for designing $\hat{H}(t)$~\cite{Wang22}.
Here we briefly review the idea of QQFT.
The Floquet Hamiltonian is defined by the fact that $e^{-i\hat{H}_F}$
is the evolution operator of quantum state over one time period.
The QQFT protocol gives a sequence of
local Hamiltonians, denoted by $\hat{H}_1,\hat{H}_2,\cdots \hat{H}_D$,
which are consecutively engineered so that the evolution
operator can be factorized as $e^{-i\hat{H}_F} = e^{-i\hat{H}_D/D} \cdots e^{-i\hat{H}_2/D}
e^{-i\hat{H}_1/D}$, where $D$ is the depth of the Hamiltonian sequence and
$1/D$ is the lifetime of each Hamiltonian. To obtain the $\hat{H}_p$s, we 
utilize the fact that $\hat{H}_F=\sum_{k} E(k)
\hat{c}^\dag_{k} \hat{c}_{k}$ is quadratic. On a lattice of size $L$,
we perform the Fourier transform $\hat{c}^\dag_{k}=\sum_{j}
\frac{e^{ikj}}{\sqrt{L}} \hat{\psi}^\dag_{j}$ with $\hat{\psi}^\dag_{j}$
being the onsite creation operator, and then reexpress the Floquet Hamiltonian
as $\hat{H}_F= \hat{\Psi}^\dag \mathcal{H}\hat{\Psi}$, where $\hat{\Psi}$
is the array of $\hat{\psi}_{j}$s and $\mathcal{H}$
is a Hermitian matrix with the elements being
\be
\label{eq:app:Hjjp}
\mathcal{H}_{j,j'} = \sum_k E(k)\frac{e^{ik\left(j-j'\right)}}{L} .
\ee
To proceed, we exploit a formula of quadratic-exponent operators,
which can be easily derived from the Baker-Campbell-Hausdorff formula. For arbitrary
Hermitian matrices $\mathcal{H}_1, \mathcal{H}_1, \cdots \mathcal{H}_d$
and a single Hermitian matrix ${\mathcal{H}}$ that satisfy
\be
e^{-i \mathcal{H}_d} \cdots 
e^{-i \mathcal{H}_2 }
e^{-i \mathcal{H}_1 } = e^{-i {\mathcal{H}} },
\ee
we always have
\be
\label{eq:app:epsiepsi}
e^{-i \hat{\Psi}^\dag \mathcal{H}_d \hat{\Psi}}\cdots 
e^{-i \hat{\Psi}^\dag \mathcal{H}_2 \hat{\Psi}}
e^{-i \hat{\Psi}^\dag \mathcal{H}_1 \hat{\Psi}} = e^{-i \hat{\Psi}^\dag
{\mathcal{H}} \hat{\Psi}}.
\ee
Equation~\eqref{eq:app:epsiepsi} simply says that the factorization
of an evolution operator with quadratic Hamiltonian (such as $e^{-i\hat{H}_F}$) is equivalent
to the factorization of the corresponding unitary matrix $e^{-i\mathcal{H}}$.
To make $\hat{H}_p = \hat{\Psi}^\dag \mathcal{H}_p \hat{\Psi}$ a local Hamiltonian,
we need to ask the $L$-by-$L$ matrix $\mathcal{H}_p$ to be local. In the QQFT protocol,
each $\mathcal{H}_p$ contains only the diagonal elements (onsite potentials)
and the off-diagonal elements $\mathcal{H}_{j,j+1}$
(hopping between nearest-neighbor sites).
Observing Eq.~\eqref{eq:app:Hjjp}, we immediately find the next factorization:
\be
\label{eq:app:eiH}
e^{-i {\mathcal{H}} } =  e^{-i e^{i \mathcal{F}} \mathcal{E} e^{ -i \mathcal{F}}} 
= e^{i \mathcal{F}} e^{-i \mathcal{E}} e^{ -i \mathcal{F}},
\ee
where $\mathcal{E} $ is a diagonal matrix with the diagonal elements being
$E(k)$, and $\mathcal{F}$ is defined by
$\left(e^{i \mathcal{F}}\right)_{j,j'}=\frac{1}{\sqrt{L}} e^{i2\pi jj'/L}$. In Eq.~\eqref{eq:app:eiH}, $\mathcal{E} $
is already diagonal and then satisfies the locality condition.
Furthermore, $e^{i \mathcal{F}}$ is recognized to be the Fourier transformation,
which can then be factorized into a sequence of local unitary matrices by
using the algorithm of quantum Fourier transform (see Ref.~[\onlinecite{Wang22}]
for the detail). The factorization of $e^{i \mathcal{F}}$ depends only upon
the value of $L$. The analytical expressions of $\mathcal{H}_p$s have been
obtained, as $L$ is an integer power of 2, i.e. $L=2^l$.
The sequence depth of $e^{i \mathcal{F}}$ scales as $L\ln L$. 

As an example, we give the sequence of Hamiltonians
that generate the required dispersion relation on a one-dimensional lattice of length $L=2^3=8$.
For simplicity, we label the lattice sites as $j=0, 1, \cdots, 7$.
In this case, the unitary evolution over a single period can be factorized into
\be\label{eq:app:L8evo}
e^{-i\mathcal{H}}= R^{(2)} A^{(2)} R^{(2)\dag}R^{(1)} A^{(1)} R^{(2)\dag}
A^{(0)}R^{(2)\dag} e^{-i \mathcal{E}} R^{(2)} A^{(0)\dag}R^{(2)} A^{(1)\dag}R^{(1)\dag}
R^{(2)} A^{(2)\dag}R^{(2)\dag}.
\ee
Here, $R^{(1)}$ and $R^{(2)}$ are the permutation matrices, which
are realized by using a sequence of swaps, say $R^{(1)} = S^{\left(1,2\right)} \ S^{\left(5,6\right)}$
and $R^{(2)} = S^{\left(3,4\right)} \ S^{\left(4,5\right)} S^{\left(5,6\right)} S^{\left(2,3\right)}
S^{\left(3,4\right)} S^{\left(1,2\right)}  $, respectively.
$S^{\left(j,j+1\right)}$ is the swap (the Pauli matrix $\sigma_x$)
between two neighbor sites $j$ and $j+1$. For the realization of $S^{\left(j,j+1\right)}$,
the corresponding Hamiltonian is $h_{j,j+1}=h_{j+1,j}=-h_{j,j}=-{h}_{j+1,j+1} = \pi/2$
and $h_{i,i'}=0$ for $i,i'\neq j,j+1$ (it is easy to verify $S^{\left(j,j+1\right)}=e^{-ih}$).
The Hamiltonian $h$ is definitely a local one, involving only an operation on two neighbor sites.
$A^{(q)}$ with $q=0,1,2$ is the local Fourier matrix, which couples $2j$ with $2j+1$ sites for
$j=0,1,2,3$. Its nonzero matrix elements are
\be
\begin{split}
\left( \begin{array}{ccc} 
A^{(q)}_{2j,2j} = \displaystyle\frac{1}{\sqrt{2}} & & A^{(q)}_{2j,2j+1}=
\displaystyle\frac{1}{\sqrt{2}}e^{i2\pi \left(j\% 2^q\right)/2^{q+1}}  \\
A^{(q)}_{2j+1,2j} =\displaystyle\frac{1}{\sqrt{2}}  & & A^{(q)}_{2j+1,2j+1}=
-\displaystyle\frac{1}{\sqrt{2}}e^{i2\pi \left(j\% 2^q\right)/2^{q+1}} \end{array} \right),
\end{split}
\ee
where $\%$ denotes the remainder. The corresponding Hamiltonian, i.e. $i \ln [A^{(q)}]$,
has only the couplings between two nearest-neighbor sites. Finally, the Hamiltonian
$\mathcal{E}$ in Eq.~\eqref{eq:app:L8evo} is made of the on-site potentials.
For a linear dispersion $E(k)=w k$, the elements of $\mathcal{E}$ can be written as $\mathcal{E}_{i,j}=\delta_{i,j}
\displaystyle\frac{2\pi}{L} j w$. One can also use the modulo $2\pi$ operation
to force $\mathcal{E}_{i,j}$ to be in the interval $[-\pi,\pi)$.
In the construction of the Hamiltonian sequence, we notice that
multiple swaps that are commutative with each other
can be combined into one without breaking the locality of Hamiltonian.
For example, $S^{\left(1,2\right)}$ and $S^{\left(5,6\right)}$ in $R^{(1)}$ can be realized
by using a single Hamiltonian that has the coupling between site-$1$ and site-$2$ and at the
same time, also the coupling between site-$5$ and site-$6$. Such a consideration
reduces the depth of the Hamiltonian sequence. In the case of $L=8$,
we find the depth to be $D=39$. The sequence consists of $32$ swaps, six $A^{(q)}$
and one $e^{-i \mathcal{E}}$.

\section{Mixing symmetry of the single-particle propagator}

In the main text, we derived from the multiplication rule that
$\ket{k',\alpha'}=\hat{U}(1,0,0)\ket{k,\alpha}$, which illustrates the transformation
of a single-particle state under $\hat{U}(1,0,0)$. In the language of many-body
physics, it is more convenient to define $\hat{U}(1,0,0)$ based on its action on the
creation or annihilation operators. This can be expressed as $\hat{c}^\dag_{k'\alpha'} = \hat{U}(1,0,0)
\hat{c}^\dag_{k\alpha}  \hat{U}^\dag (1,0,0)$. The field operators in real space
are obtained through Fourier transformation of $\hat{c}^\dag_{k\alpha} $, given by
$\hat{\psi}^\dag_{x\alpha} = \sum_k 
\displaystyle\frac{e^{-ikx}}{\sqrt{L}} \hat{c}^\dag_{k\alpha}$, where $L$ is the system size.
The time evolution of field operators is defined as $\hat{\psi}^\dag_{x \alpha}(t)
= e^{i\hat{H}_F t} \hat{\psi}^\dag_{x \alpha} e^{-i\hat{H}_F t}$ for integer $t$
(integer multiples of the period). Utilizing Eq.~\eqref{eq:app:ekpk}, we can derive
the following expression:
\be\label{eq:app:ufiop}
\hat{U}(1,0,0) \hat{\psi}^\dag_{x \alpha}(t) \hat{U}^\dag (1,0,0) =
 \hat{\psi}^\dag_{x' \alpha'}(t'),
\ee
where $\left(t',x'\right)^T = A\left(t,x\right)^T$, and $t,x,t',x'$ are all integers.
The transformation $\hat{U}(1,0,0)$ induces changes in
both the spatial and temporal coordinates of the field operators,
which are determined by the matrix $A$.

The propagator of particles in band-$\alpha$ is defined as
\be\label{eq:app:gt1x1}
G_\alpha(t_1 x_1,t_2x_2)=-i \theta(t_1-t_2) \langle \left[\hat{\psi}_{x_1\alpha}(t_1), 
\hat{\psi}^\dag_{x_2\alpha}(t_2)\right]_{\pm} \rangle,
\ee
where the plus (minus) sign corresponds to fermions (bosons), and $\theta$
represents the Heaviside function. The coordinates $t_1,x_1,t_2,x_2$ are all integers.
The angle brackets $\langle\rangle$ denote the expectation
value with respect to the vacuum state. Due to discrete translational
symmetry, $G_\alpha$ depends only on the difference $\Delta t=t_1-t_2$ and
$\Delta x=x_1-x_2$ for integer coordinates. Using Eq.~\eqref{eq:app:ufiop},
we immediately find:
\be\label{eq:app:galgal}
G_\alpha(\Delta t,\Delta x) = G_{\alpha'}(\Delta t',\Delta x'),
\ee
where $(\Delta t',\Delta x')^T=A (\Delta t,\Delta x)^T$. This equation explains how
the mixing symmetry manifests in the particle propagator. For $\alpha'=\alpha$
(a singlet band in the $\mathcal{P}_2$ class), the propagator must remain invariant
after a linear operation $A$ on the spacetime coordinates, imposing
a strong constraint on the propagator. For $\alpha'\neq \alpha$,
the propagator of band-$\alpha$ after the coordinate transformation
becomes the propagator of band-$\alpha'$. Thus, Eq.~\eqref{eq:app:galgal}
establishes a connection between propagators of different bands.

In experiments, what can be measured is the wave function, or more precisely,
the absolute magnitude of the wave function. The wave function is directly linked
to the propagator. If we initially locate a particle at position $x=0$ at time $t=0$,
its wave function at a later time satisfies, according to
Eq.~\eqref{eq:app:gt1x1} and~\eqref{eq:app:galgal},
\be
\Psi_\alpha(t,x) = \Psi_{\alpha'} \left( t',x'\right),
\ee
where $\left( t',x'\right)^T = A \left( t,x\right)^T$ and $t,x$ are arbitrary integers.
An alternative way to prove this result is by using
$\Psi_\alpha (t,x) = \sum_k {e^{ikx-itE_\alpha(k)}}/{L}$ and Eq.~\eqref{eq:app:ekpk}.

\end{widetext}

\end{document}